\documentclass[graybox]{svmult}

\usepackage{graphics} % for pdf, bitmapped graphics files
\usepackage{amsmath} % assumes amsmath package installed
\usepackage{amssymb}  % assumes amsmath package installed

\usepackage[latin1]{inputenc}
\usepackage[english]{babel}
\usepackage{graphicx}
\usepackage{amsmath}
\usepackage{amssymb}
\usepackage{color}
\usepackage{subfigure}

\makeindex

\begin{document}

\title*{Systematic and multifactor risk models revisited}
\author{Michel Fliess, Cédric Join}
\institute{Michel Fliess\at LIX (CNRS, UMR 7161),\\ \'Ecole polytechnique,\\ 91128 Palaiseau, France.\\
\email{Michel.Fliess@polytechnique.edu}
\and 
C\'edric Join \at Non-A -- INRIA \& CRAN (CNRS, UMR 7039),\\ Universit\'e de Lorraine,\\ BP 239,\\ 54506 Vandoeuvre-l\`es-Nancy, France.\\ \email{cedric.join@univ-lorraine.fr}
\and
M.F. \& C.J. \at AL.I.E.N. (ALgèbre pour Identification \& Estimation Numériques) S.A.S., \\ 24-30 rue Lionnois, BP 60120, 54003 Nancy, France. \\
\email{\{michel.fliess, cedric.join\}@alien-sas.com}}
\maketitle

\abstract{Systematic and
multifactor risk models are revisited via methods which were already
successfully developed in signal processing and in automatic
control. The results, which bypass the usual criticisms on those
risk modeling, are illustrated by several  successful computer experiments.}

\newpage

\section{Introduction}
\emph{Systematic}, or \emph{market}, \emph{risk} is one of the most
studied risk models not only in financial engineering, but also in actuarial sciences, in business and corporate management, and in 
several other domains. It is associated to the beta ($\beta$)
coefficient, which is familiar in the investment industry since
Sharpe's {\em capital asset pricing model} (\emph{CAPM})
\cite{sharpe1}. The pitfalls and shortcomings of $\beta$ have been
detailed by a number of excellent authors.\footnote{The literature
questioning the validity of the beta coefficient is huge and well
summarized in several textbooks (see, \textit{e.g.}, \cite{bodie}).
A recent and remarkable paper by Tofallis \cite{tofa} has been most
helpful in this study.} Replacing moreover time-invariant linear
regressions by time-varying and/or nonlinear ones does not seem to
improve this situation.\footnote{See, {\it e.g.}, \cite{adrian,ter},
and the references therein.} The \emph{model-free} standpoint
advocated in \cite{beta} and \cite{douai} alleviates several of the
known deficiencies but unfortunately cannot be extended to
multifactor risk models which became also popular after Ross'
\emph{arbitrage pricing theory} (\emph{APT}) \cite{ross}. In order
to encompass the univariate and multivariate cases, we propose here
a unified definition, with the same advantages, namely a clear-cut
mathematical foundation, which
\begin{itemize}
\item bypasses clumsy statistical and/or financial
assumptions,
\item leads to efficient computations.
\end{itemize}
Our approach is based on the following ingredients:
\begin{itemize}
\item As in our previous works \cite{fes,beta,douai,agadir} we utilize the
Cartier-Perrin theorem \cite{cartier}. It shows that under a mild
integrability condition any time series may decomposed as a sum of a
\emph{mean}, or \emph{trend}, and of \emph{quick fluctuations}.
\item Classic mathematical tools like the Wronski determinants
\cite{ince}.
\item We employ recent estimation and identification
techniques\footnote{The use of advanced theory stemming from signal
analysis is not new in finance. See, \textit{e.g.}, \cite{gen}.}
\cite{sira1,sira2}, which are stemming from control theory and
signal processing where they have been utilized quite
successfully.\footnote{See, \textit{e.g.},
\cite{bruit,shannon,abrupt,diag,nl,mexico,mboup1,mboup2,trap1,trap2,trap3,ushi},
and the references therein.}
\end{itemize}
From a more practical standpoint, our main result is the
derivation of two independent $\beta$ coefficients, the first one
for the comparison between returns and the second one for the
comparison between volatilities. It implies among other consequences
that the importance of the popular $\alpha$ coefficient might
vanish.

Our paper is organized as follows. After a short review of the
Cartier-Perrin theorem, Section \ref{theory} details the new
mathematical definitions of the coefficients $\alpha$ and $\beta$,
and of $\beta$ alone. Section \ref{comment} develops the comparison
with the classical settings. Computer illustrations are provided in
Section \ref{computer}.

Future publications will be exploiting the above advances in at
least three directions:
\begin{enumerate}
\item The extension of Section \ref{vola} to skewness and kurtosis
should be straightforward. Our understanding, which would not rely exclusively
any more on ``Gaussianism'', of the respective behaviors of various
assets might therefore be quite enhanced.
\item According to the methods
sketched in \cite{agadir} and in \cite{pricing}, dynamic portfolio
management and option pricing may be achieved by tracking quite
independent performances with respect to returns and volatilities.
\item We will relate some instances of \emph{systemic risk} to the \emph{abrupt changes}
\cite{abrupt} of some quantities like our new beta coefficients (see
\cite{fes,beta,douai} for preliminary results).
\end{enumerate}

\section{Theoretical background}\label{theory}
\subsection{A short review on time series via nonstandard analysis}
Take the time interval $[0, 1] \subset \mathbb{R}$ and introduce as
often in \emph{nonstandard analysis} the infinitesimal sampling
$${\mathfrak{T}} = \{ 0 = t_0 < t_1 < \dots < t_N = 1 \}$$
where $t_{\iota + 1} - t_{\iota}$, $0 \leq \iota < N$, is {\em
infinitesimal}, {\it i.e.}, ``very small''.\footnote{See,
\textit{e.g.}, \cite{diener1,diener2} for basics in nonstandard
analysis.} A \emph{time series} $X(t)$ is a function $X:
{\mathfrak{T}} \rightarrow \mathbb{R}$.

The {\em Lebesgue measure} on ${\mathfrak{T}}$ is the function
$\ell$ defined on ${{\mathfrak{T}}} \backslash \{1\}$ by
$\ell(t_{i}) = t_{i+1} - t_{i}$. The measure of any interval $[c, d]
\subset \mathfrak{I}$, $c \leq d$, is its length $d -c$.  The
\emph{integral} over $[c, d]$ of the time series $X(t)$ is the sum
$$\int_{[c, d]} Xd\tau = \sum_{t \in [c, d]} X(t)\ell(t)$$
$X$ is said to be $S$-{\em integrable} if, and only if, for any
interval $[c, d]$ the integral $\int_{[c, d]} |X| d\tau$ is
\emph{limited}, \textit{i.e.} not infinitely large, and, if $d - c$
is infinitesimal, also infinitesimal.

$X$ is $S$-{\em continuous} at $t_\iota \in {\mathfrak{T}}$ if, and
only if, $f(t_\iota) \simeq f(\tau)$ when $t_\iota \simeq
\tau$.\footnote{$a \simeq b$ means that $a - b$ is infinitesimal.}
$X$ is said to be {\em almost continuous} if, and only if, it is
$S$-continuous on ${\mathfrak{T}} \setminus R$, where $R$ is a {\em
rare} subset.\footnote{The set $R$ is said to be \emph{rare}
\cite{cartier} if, for any standard real number $\alpha > 0$, there
exists an internal set $B \supset A$ such that $m(B) \leq \alpha$.}
$X$ is \emph{Lebesgue integrable} if, and only if, it is
$S$-integrable and almost continuous.

A time series ${\mathcal{X}}: {\mathfrak{T}} \rightarrow \mathbb{R}$
is said to be {\em quickly fluctuating}, or {\em oscillating}, if,
and only if, it is $S$-integrable and $\int_A {\mathcal{X}} d\tau$
is infinitesimal for any {\em quadrable} subset.\footnote{A set is
\emph{quadrable} \cite{cartier} if its boundary is rare.}

Let $X: {\mathfrak{T}} \rightarrow \mathbb{R}$ be a $S$-integrable
time series. Then, according to the Cartier-Perrin theorem
\cite{cartier},\footnote{See \cite{lobry} for a more down to earth
exposition.} the additive decomposition
\begin{equation}\label{decomposition}
\boxed{X(t) = E(X)(t) + X_{\tiny{\rm fluctuat}}(t)}
\end{equation}
holds where
\begin{itemize}
\item the \emph{mean} $E(X)(t)$ is Lebesgue integrable,
\item $X_{\tiny{\rm fluctuat}}(t)$ is quickly fluctuating.
\end{itemize}
The decomposition \eqref{decomposition} is unique up to an
infinitesimal.

\begin{remark}Decomposition \eqref{decomposition}, where $E(X)(t)$
is ``smoother'' than $X(t)$, provides, to the
best of our knowledge, the first complete theoretical justification
(see \cite{fes}) of the \emph{trends} in \emph{technical analysis}
(see, \textit{e.g.}, \cite{bechu,kirk}).
\end{remark}

\subsection{Multivariate factors}
\subsubsection{Arithmetical average}
Assume that $X: {\mathfrak{T}} \rightarrow \mathbb{R}$ is
$S$-integrable. Take a quadrable set $A \subseteq {\mathfrak{T}}$
such that $\int_A d\tau$ is \emph{appreciable}, \textit{i.e.},
non-infinitesimal. The \emph{arithmetical average} of $X$ on $A$,
which is written $\text{AV}_A (X)$, is defined by
\begin{equation*}
\boxed{\text{AV}_A (X) = \frac{\int_A X d\tau}{\int_A d\tau}}
\end{equation*}
It follows at once from Equation \eqref{decomposition} that the
difference between $\text{AV}_A (X)$ and $\text{AV}_A (E(X))$ is
infinitesimal, \textit{i.e.},
$$
\text{AV}_A (X) \simeq \text{AV}_A (E(X))
$$
In practice, $A$ is a time interval $[t-L, t]$, with an appreciable
length $L$. Set, if $t \geq L$,
{%\small 
\begin{equation}\label{aver}
\boxed{\overline{X} (L, t) = \text{AV}_{[t- L, t]} (X) =
\frac{\int_{t-L}^{t} X d\tau}{L} \simeq \frac{\int_{t-L}^{t} E(X)
d\tau}{L}}
\end{equation}
Introduce
\begin{equation}\label{cauchy}
\overline{X}^{[\nu]} (L, t) = \frac{1}{(\nu - 1)!} \int_{t-L}^{t} (t
- \tau)^{\nu - 1} X d\tau
\end{equation}}
It corresponds via the classic Cauchy formula to an iterated
integral of order $\nu$ (see, \textit{e.g.}, \cite{folland}). Note
that $\overline{X}^{[1]}  (L, t) = \overline{X}  (L, t)$.

\subsection{Alpha and betas}\label{ab}
Take $n+1$ $S$-integrable time series $Y, X_1, \dots, X_n:
{\mathfrak{T}} \rightarrow \mathbb{R}$. Assume, without any loss of
generality, that their values at any $t_\iota \in {\mathfrak{T}}$ is
bounded by a given limited number. Set
{\footnotesize \begin{equation*}\label{alphabeta}
\boxed{\overline{Y}  (L, t) = \alpha(L, t) + \beta_1 (L,t)
\overline{X}_1 (L, t) + \dots + \beta_n (L,t) \overline{X}_n (L, t)}
\end{equation*}}
\noindent $\alpha(L, t)$, $\beta_i (L,t) \in \mathbb{R}$, $i = 1, \dots,
n$, are not yet uniquely determined.

Define the time series $\mathbf{1}: {\mathfrak{T}} \rightarrow
\mathbb{R}$, $t_\iota \mapsto 1$. Its arithmetical average is always
$1$. Equation \eqref{cauchy} yields
\begin{equation*}\label{11}
\overline{\mathbf{1}}^{[\nu]} (L, t) = \frac{L^{\nu - 1}}{\nu!}
\end{equation*}
Introduce the \emph{Wronskian}-like determinant (see, \textit{e.g.},
\cite{ince})
{\small \begin{equation}\label{w}
W_{{\mathbf{1}}, X_1, \dots, X_n} (L, t) =
\begin{vmatrix}1 & \overline{X}^{[1]}_1 (L,t)
& \dots & \overline{X}^{[1]}_n (L, t) \\
\dots & \dots & \dots & \dots \\
\frac{L^{n}}{(n + 1)!} & \overline{X}^{[n + 1]}_1 (L,t) & \dots &
\overline{X}^{[n + 1]}_n (L, t)
\end{vmatrix}
\end{equation}}
$X_1, \dots, X_n$ are said to be \emph{$\alpha$-W-independent} on
$[t-L, t]$ if, and only if, $W_{{\mathbf{1}}, X_1, \dots, X_n} (L,
t)$ is appreciable.

Introduce the $(n + 1) \times (n + 2)$ matrix
{\small \begin{equation}\label{n+2}
{\mathcal{M}}_{Y, {\mathbf{1}}, X_1, \dots, X_n} (L, t) =
\begin{pmatrix}
\overline{Y}^{[1]} (L, t) & 1 & \overline{X}^{[1]}_1
(L, t) & \dots & \overline{X}^{[1]}_n (L, t) \\
\dots & \dots & \dots & \dots & \dots \\
\overline{Y}^{[n + 1]} (L, t) & \frac{L^{n}}{(n + 1)!} &
\overline{X}^{[n + 1]}_1 (L, t) & \dots & \overline{X}^{[n + 1]}_n
(L, t)
\end{pmatrix}
\end{equation}}
Assume that $X_1, \dots, X_n$ are $\alpha$-W-independent on $[t-L,
t]$. Then the matrix \eqref{n+2} is of rank $n + 1$. The Cramer rule
yields limited values for $\alpha (L, t)$, $\beta_1 (L,t)$, \dots,
$\beta_n (L,t)$ in Equation \eqref{alphabeta}:
{\small $$
\alpha (L, t) = \frac{\begin{vmatrix}\overline{Y}^{[1]} (L,t) &
\overline{X}^{[1]}_1 (L,t)
& \dots & \overline{X}^{[1]}_n (L, t) \\
\dots & \dots & \dots & \dots \\
\overline{Y}^{[n + 1]} (L,t) & \overline{X}^{[n + 1]}_1 (L,t) &
\dots & \overline{X}^{[n + 1]}_n (L, t)
\end{vmatrix}}{W_{{\mathbf{1}}, X_1, \dots, X_n} (L, t)}
$$}
{\small $$
\beta_1 (L, t) = \frac{\begin{vmatrix}1 & \overline{Y}^{[1]} (L,t) &
\overline{X}^{[1]}_2 (L, t)
& \dots & \overline{X}^{[1]}_n (L, t) \\
\dots & \dots & \dots & \dots \\
\frac{L^{n}}{(n + 1)!} & \overline{Y}^{[n + 1]} (L,t) &
\overline{X}^{[n+1]}_2 (L, t) & \dots & \overline{X}^{[n + 1]}_n (L,
t)
\end{vmatrix}}{W_{{\mathbf{1}}, X_1, \dots, X_n} (L, t)}
$$}
\begin{center}
\dots
\end{center}
{\small $$
\beta_n (L, t) = \frac{\begin{vmatrix}1 & \overline{X}^{[1]}_1 (L,t)
& \dots & \overline{X}^{[1]}_{n-1} (L,t) & \overline{Y}^{[1]} (L, t) \\
\dots & \dots & \dots & \dots \\
\frac{L^{n}}{(n + 1)!} & \overline{X}^{[n + 1]}_1 (L,t) & \dots &
\overline{X}^{[n + 1]}_{n-1} (L,t) & \overline{Y}^{[n + 1]} (L, t)
\end{vmatrix}}{W_{{\mathbf{1}}, X_1, \dots, X_n} (L, t)}
$$}
\begin{remark}\label{rem}
Replacing in Equation \eqref{alphabeta} the arithmetic averages by
the original time series yields
$$
Y = \alpha(L, t) + \beta_1 (L,t) X_1 + \dots + \beta_n (L,t) X_n +
e_{[t-L, t]}
$$
where $\int_{t-L}^{t} e_{[t-L, t]} d\tau$ is infinitesimal.
\end{remark}

\subsection{Betas alone}\label{bet}
Let us drop $\alpha$. Equation \eqref{alphabeta} becomes
\begin{equation}\label{betas}
\overline{Y}  (L, t) = \sum_{i=1}^{n} \beta_i (L,t) \overline{X}_i
(L, t)
\end{equation}
Determinant \eqref{w} is replaced by
{\small $$
W_{X_1, \dots, X_n} (L, t) = \begin{vmatrix}\overline{X}^{[1]}_1
(L,t)
& \dots & \overline{X}^{[1]}_n (L, t) \\
\dots & \dots & \dots \\
\overline{X}^{[n]}_1 (L,t) & \dots & \overline{X}^{[n]}_n (L, t)
\end{vmatrix}
$$}\noindent $X_1, \dots, X_n$ are said to be \emph{W-independent} on $[t-L, t]$
if, and only if, $W_{X_1, \dots, X_n} (L, t)$ is appreciable. Matrix
\eqref{n+2} is replaced by the $n \times (n+1)$ matrix
{\small $$
{\mathcal{M}}_{Y, X_1, \dots, X_n} (L, t) =
\begin{pmatrix}
\overline{Y}^{[1]} (L, t) & \overline{X}^{[1]}_1
(L, t) & \dots & \overline{X}^{[1]}_n (L, t) \\
\dots & \dots & \dots & \dots \\
\overline{Y}^{[n]} (L, t) & \overline{X}^{[n]}_1 (L, t) & \dots &
\overline{X}^{[n]}_n (L, t)
\end{pmatrix}
$$}
Assume that $X_1, \dots, X_n$ are W-independent on $[t-L, t]$. Then
${\mathcal{M}}_{Y, X_1, \dots, X_n} (L, t)$ is of rank $n$. Limited
values for $\beta_i (L,t)$, $i = 1, \dots, n$ are again given by the
Cramer rule. Although we do not give again the formulae, it goes
without saying that these numerical values are in general different
from those derived in Sect. \ref{ab}. We do not repeat also Remark
\ref{rem}.
\begin{remark}\label{n1}
If $n = 1$ and $\int_{t-L}^{t} \overline{Y}d\tau \neq 0$,
\begin{equation}\label{1}
\beta_1 (L,t) = \frac{\overline{X}_1 (L, t)}{\overline{Y} (L, t)} =
\frac{\int_{t-L}^{t}\overline{X}_1 d\tau}{\int_{t-L}^{t}
\overline{Y} d\tau}
\end{equation}
\end{remark}

\section{Comments}\label{comment}
\subsection{The model-free standpoint}\label{mfc}
The length $L$ of the time window $[t-L, t]$ may be chosen quite
short, \textit{i.e.}, of a size compatible with what is needed for
calculating the trends in \cite{fes}. Updating the various factors
is achieved by letting slide this time window. Let us emphasize that
the linearity of the local models \eqref{alphabeta} and
\eqref{betas}, which are valid only during a short time interval,
does not imply therefore a global time-invariant linearity as
assumed in the CAPM and APT settings. This \emph{model-free}
standpoint has already been proved to be quite efficient in control
theory.\footnote{See \cite{ijc13}. Many successful concrete
engineering applications may be found in the references.}
\begin{remark}
Equations \eqref{alphabeta} and \eqref{betas} should not be viewed
as time-varying linear relations, since the values of their
coefficients depend nonlinearly on $X_1$, \dots, $X_n$.
\end{remark}

\subsection{Reverse formula}
Take for simplicity's sake $n = 1$ in Equation \eqref{alphabeta}.
Then
$$
\overline{Y}  (L, t) = \alpha + \beta_1 (L,t) \overline{X}_1 (L, t)
$$
yields, if $\beta_1 (L,t) \neq 0$,
$$
\overline{X}_1  (L, t) = - \frac{1}{\alpha (L,t)} + \frac{1}{\beta_1
(L,t)} \overline{Y} (L, t)
$$
The same reverse formula would have also been derived from the
linear algebra of Section \ref{bet}.

Now, we restrict ourselves for simplicity's sake to a CAPM-like
equation
\begin{equation}\label{capm}
r(t) = \alpha + \beta R (t) + \epsilon (t)
\end{equation}
where
\begin{itemize}
\item $r(t)$ and $R(t)$ are the values at time $t$ of some returns,
\item $\epsilon (t)$ is a zero-mean stochastic processes,
\item $\alpha$ and $\beta$ are constant.
\end{itemize}
As pointed out in \cite{tofa}, the classic least square techniques
utilized with Equation \eqref{capm} do not lead to the most natural reverse formula.

\subsection{Volatility}\label{vola}
\subsubsection{Today's situation}\label{today}
Consider again Equation \eqref{capm}. This global linear
time-invariant equation leads to usual {\em systematic}, or
\emph{market}, risk calculation, \textit{i.e}, to
\begin{equation}\label{risk}
\text{var}(r) = \beta^2 \text{var} (R) + \text{var} (\epsilon)
\end{equation}
where $\text{var} (\epsilon)$ should be ``small'' if there is a
``good'' diversification. It explains
\begin{enumerate}
\item why increasing $\beta$ also increases the risk,
\item the importance of generating a ``good'' $\alpha$ in Equation \eqref{capm}.
\end{enumerate}
If, as emphasized in \cite{tofa}, Equation \eqref{capm} does not
hold, \textit{i.e.}, there is no global linear time-invariant
relationship, Equation \eqref{risk} is then erroneous. The whole
``philosophy'' which was built in order to justify the utilization
of the CAPM and of its extensions like the APT (see, \textit{e.g.},
\cite{bodie}) might therefore break down.\footnote{See also the
harsh quotations and comments in \cite{bernstein}.}
\begin{remark}
Equation \eqref{aver} shows that the quick fluctuations do not
appear in Equations \eqref{alphabeta} and \eqref{betas}. Those
equations are therefore useful for comparing the time evolution of
means, \textit{i.e.}, trends, and certainly not for the comparison
of the corresponding volatilities.
\end{remark}

\subsubsection{A remedy}\label{remedy}
We start by reviewing the definitions of (co)variances and
volatility given in \cite{douai,agadir}. Take two $S$-integrable
time series $X$, $Y$ such that their squares and the squares of
$E(X)$ and $E(Y)$ are also $S$-integrable. Then the following
property is obvious:  $X Y$, $E(X) E(Y)$, $E(X)Y_{\tiny{\rm
fluctuat}}$, $X_{\tiny{\rm fluctuat}}E(Y)$, $X_{\tiny{\rm fluctuat}}
Y_{\tiny{\rm fluctuat}}$ are all $S$-integrable. Assume moreover
that $E(X)$ and $E(Y)$ are \emph{differentiable} in the following
sense: there exist two Lebesgue integrable time series $f, g:
{\mathfrak{T}} \rightarrow \mathbb{R}$, such that, $\forall ~ t \in
{\mathfrak{T}}$, with the possible exception of a limited number of
values of $t$, $E(X) = E(X)(0) + \int_{0}^{t} f(\tau)d\tau$, $E(Y) =
E(Y)(0) + \int_{0}^{t} g(\tau)d\tau$. Integrating by parts shows
that the pro\-ducts $E(X)Y_{\tiny{\rm fluctuat}}$ and $X_{\tiny{\rm
fluctuat}}E(Y)$ are quickly fluctuating.

\begin{remark}
Let us emphasize that the product $X_{\tiny{\rm
fluctuat}}Y_{\tiny{\rm fluctuat}}$ is not necessarily quickly
fluctuating.
\end{remark}

The following definitions are natural:
\begin{enumerate}
\item The \emph{covariance} of two time series $X$ and $Y$ is
%$$
\begin{eqnarray*}
%\boxed{
\mbox{\rm cov}(XY) & = & E\left((X - E(X))(Y - E(Y)) \right)
\\ & \simeq & E(XY) - E(X)(t) \times E(Y)
%}
\end{eqnarray*}
%$$
\item The \emph{variance} of the time series $X$ is
\begin{eqnarray*}\label{var}
\mbox{\rm var}(X) & = & E\left((X - E(X))^2 \right) \\ & \simeq &
E(X^2) - \left(E(X)\right)^2
\end{eqnarray*}
\item The \emph{volatility} of $X$ is the corresponding standard
deviation
\begin{equation}\label{vol}
\boxed{\mbox{\rm vol}(X) = \sqrt{\mbox{\rm var}(X)}}
\end{equation}
\end{enumerate}

The definition of volatility given by Equation \eqref{vol}
associates to a time series $X$ another time series $\text{vol}
(X)$, which is called the \emph{volatility time series}. Take now $n
+ 1$ time series $Y$, $X_1$, \dots, $X_n$, which satisfy the above
assumptions on integrability and differentiability. We may repeat
for the $n + 1$ time series $\text{vol} (Y)$, $\text{vol} (X_1)$,
\dots, $\text{vol} (X_n)$ the same calculations as in Sections
\ref{ab} and \ref{bet}. It yields new relations between those
volatilities.

\begin{remark}
Take $n = 1$ as in the CAPM setting. We now have two time-varying
betas for comparing the two assets:
\begin{enumerate}
\item The first one, derived from Sections \ref{ab} or \ref{bet},
compares an averaged time evolution of their values or returns.
\item The second one, derived from Section \ref{remedy}, compares an
averaged time evolution of their corresponding
volatilities.\footnote{Let us stress that the ``famous'' $\alpha$
coefficient, in Equation \eqref{capm}, of the CAPM might therefore
become quite obsolete.}
\end{enumerate}
\end{remark}

\section{Some computer experiments}\label{computer}
\subsection{Monovariate $\beta$}
\begin{figure}
\vspace{-.4cm}
\center
\subfigure[S\&P 500]{\rotatebox{-0}{\includegraphics*[width=.32\columnwidth]{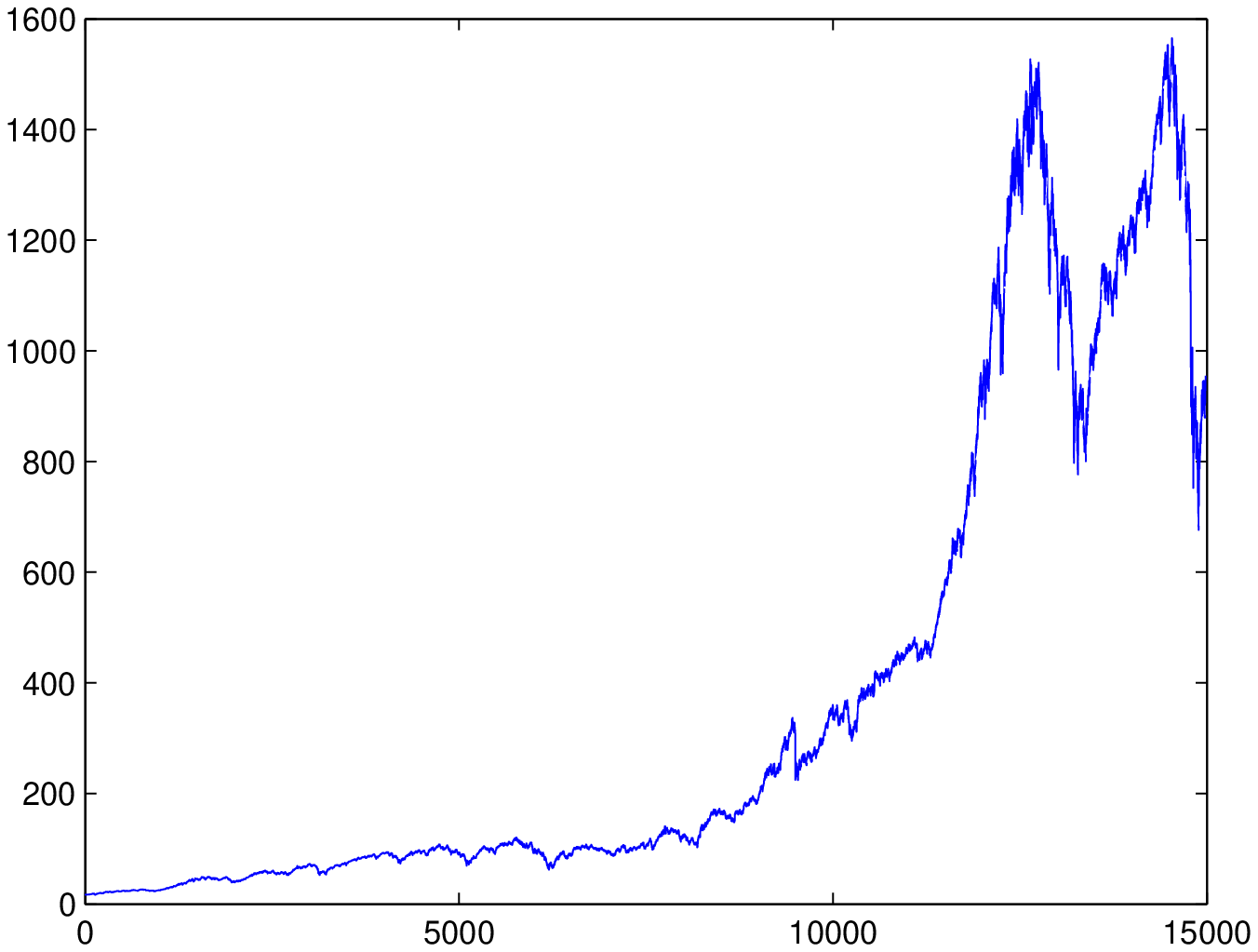}}}
\subfigure[IBM]{\rotatebox{-0}{\includegraphics*[width=.32\columnwidth]{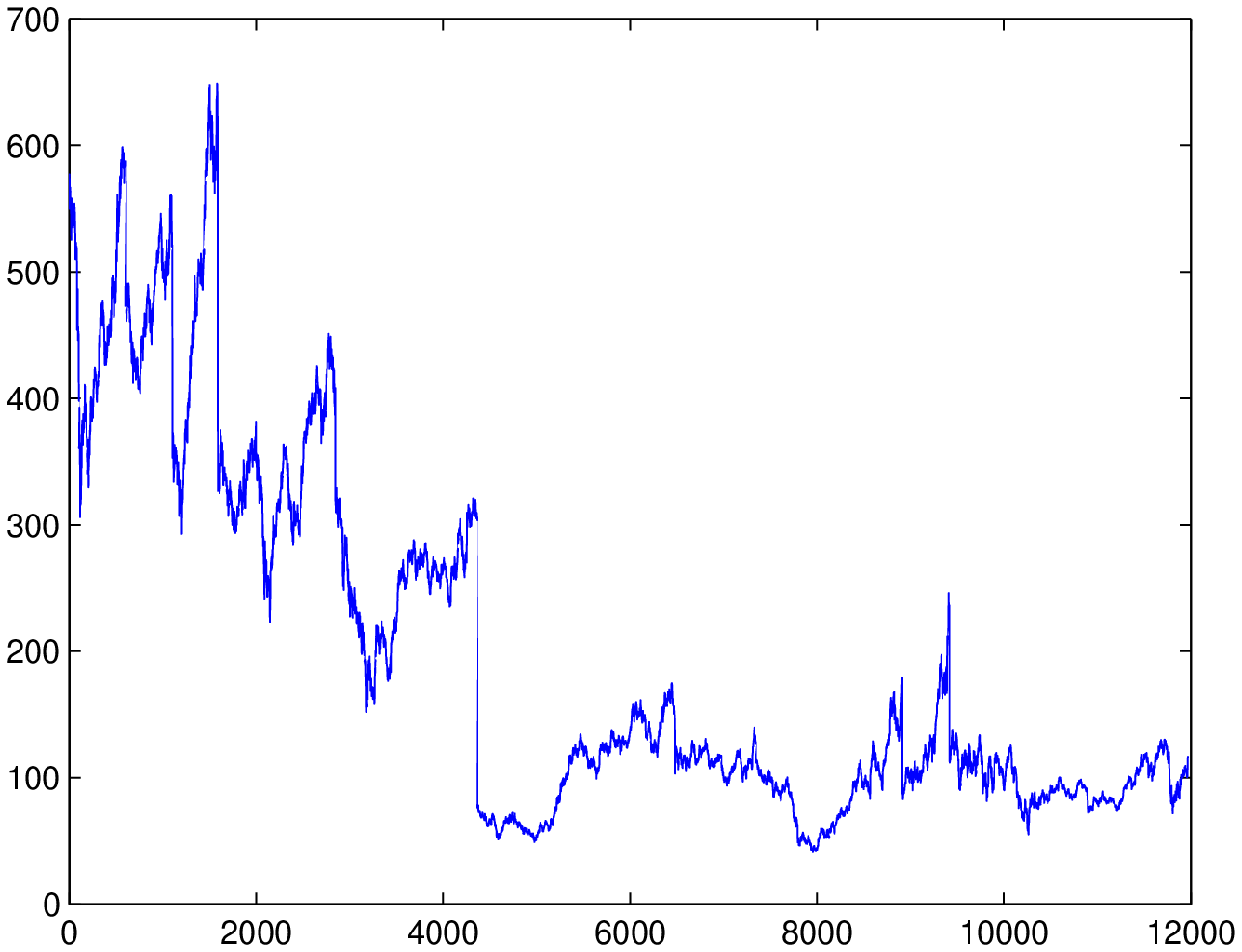}}}
\subfigure[JPM]{\rotatebox{-0}{\includegraphics*[width=.32\columnwidth]{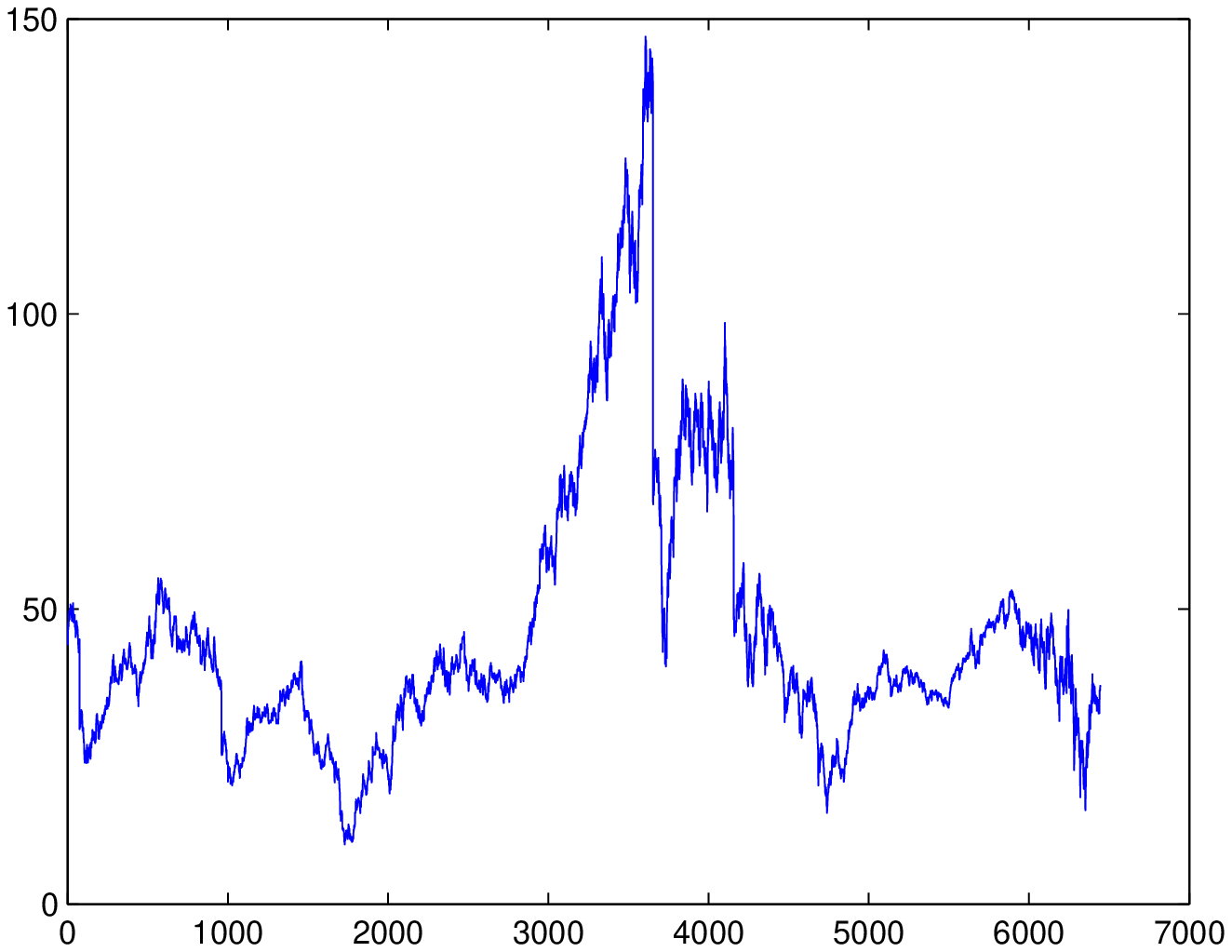}}}
\vspace{-.4cm} \caption{Time values\label{Value}}
\end{figure}

\begin{figure}
\vspace{-.4cm}
\center
\subfigure[S\&P 500]{\rotatebox{-0}{\includegraphics*[width=.32\columnwidth]{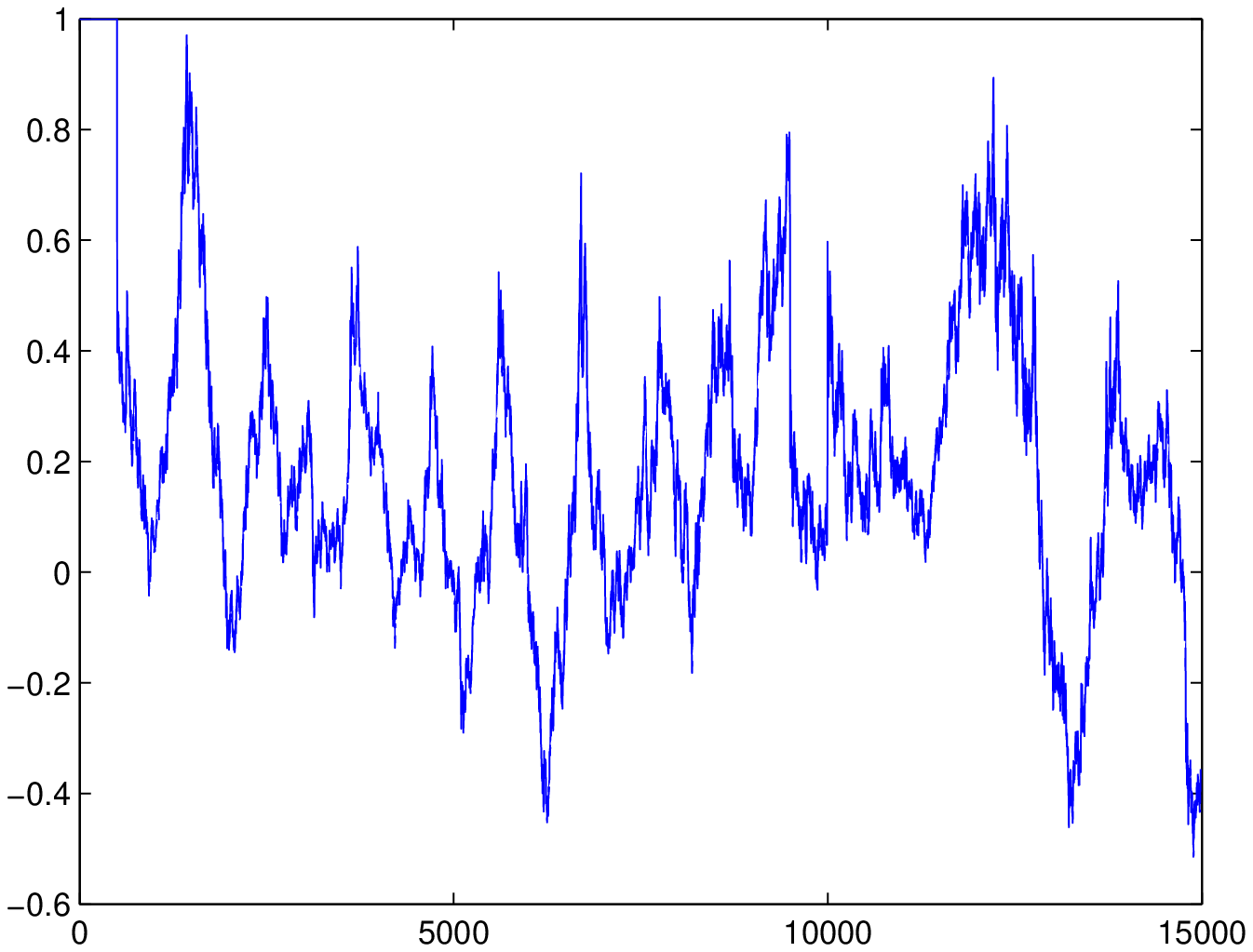}}}
\subfigure[IBM]{\rotatebox{-0}{\includegraphics*[width=.32\columnwidth]{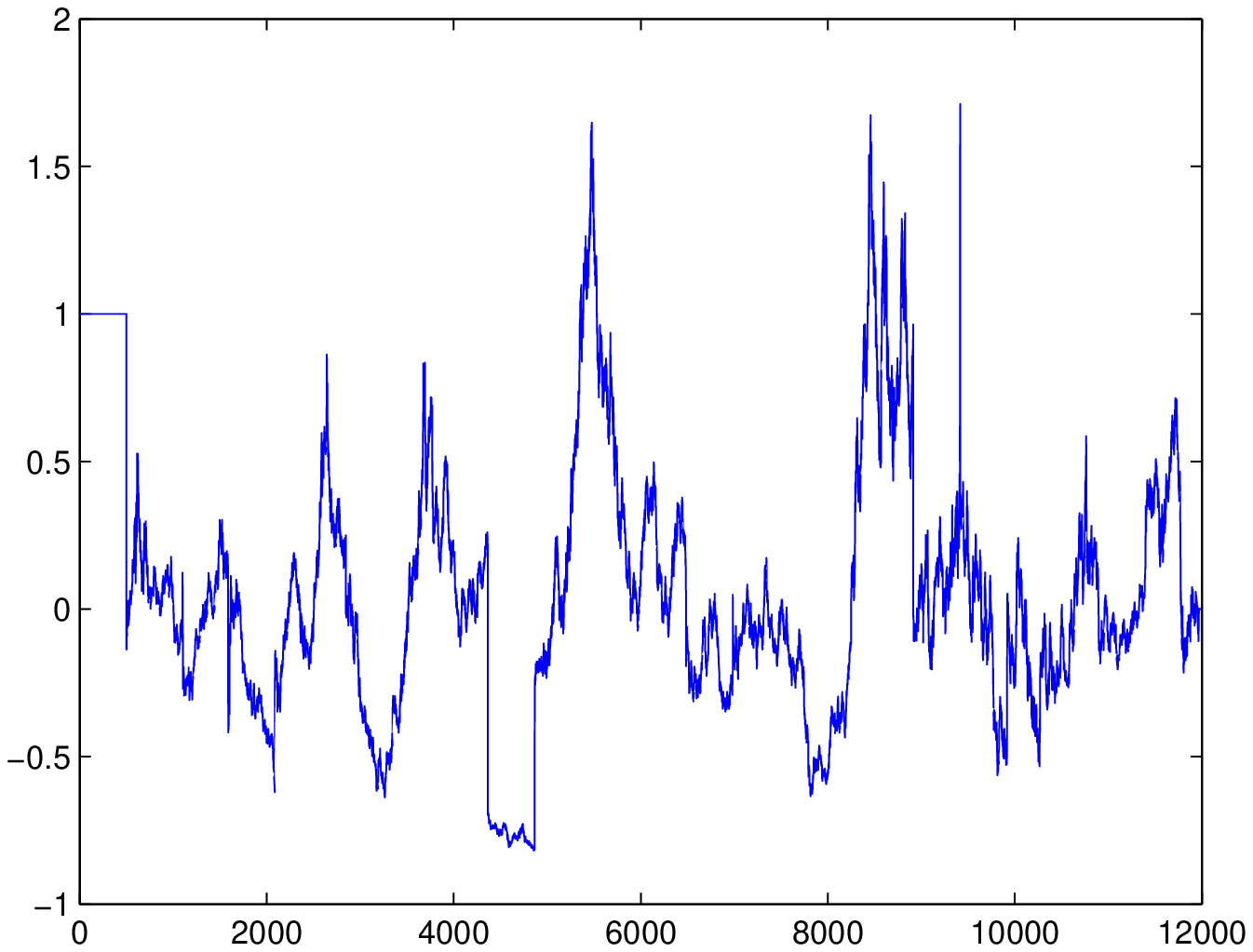}}}
\subfigure[JPM]{\rotatebox{-0}{\includegraphics*[width=.32\columnwidth]{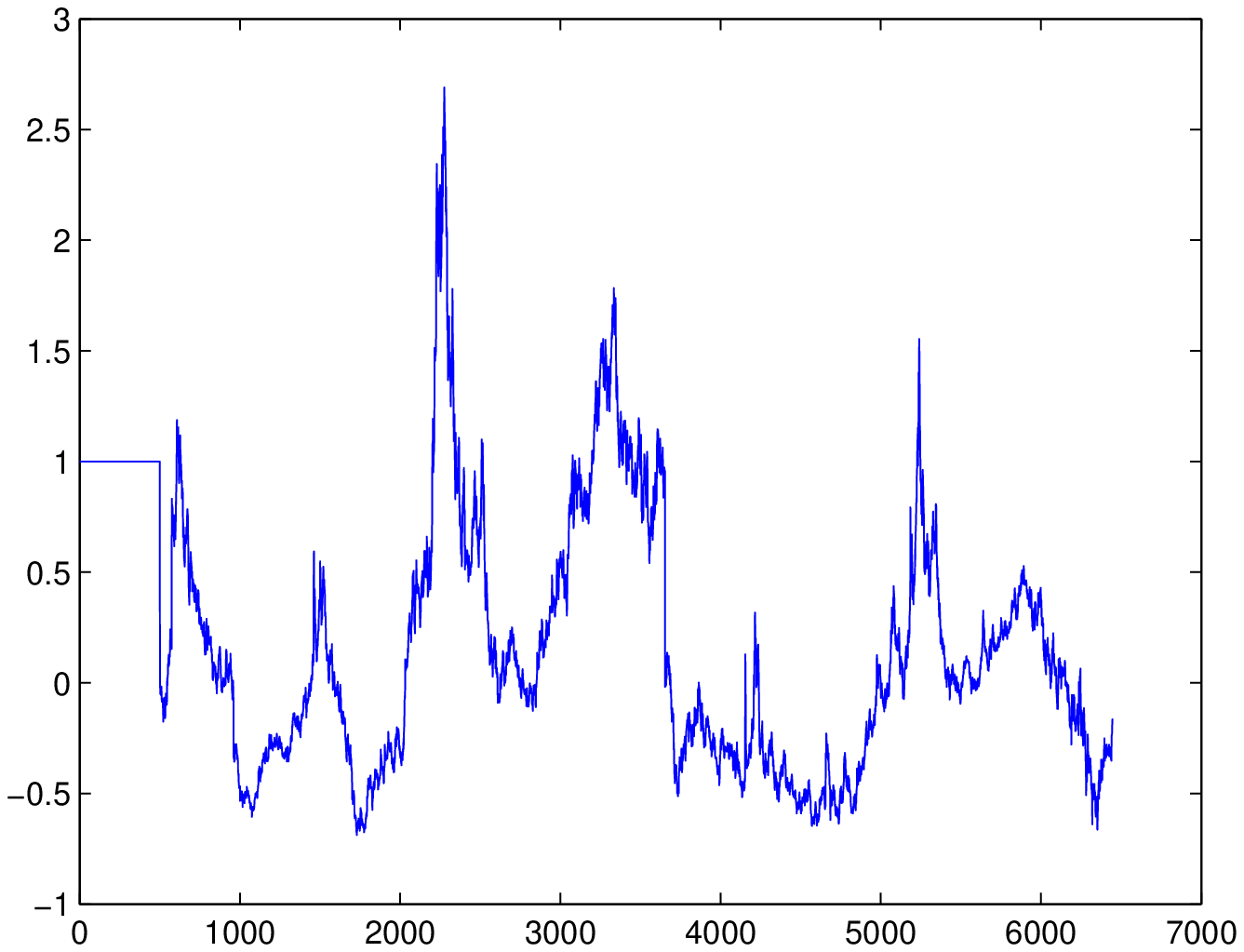}}}
\vspace{-.4cm}\caption{Returns\label{Return}}
\end{figure}

\begin{figure}
\vspace{-.4cm}
\center
\subfigure[S\&P 500]{\rotatebox{-0}{\includegraphics*[width=.32\columnwidth]{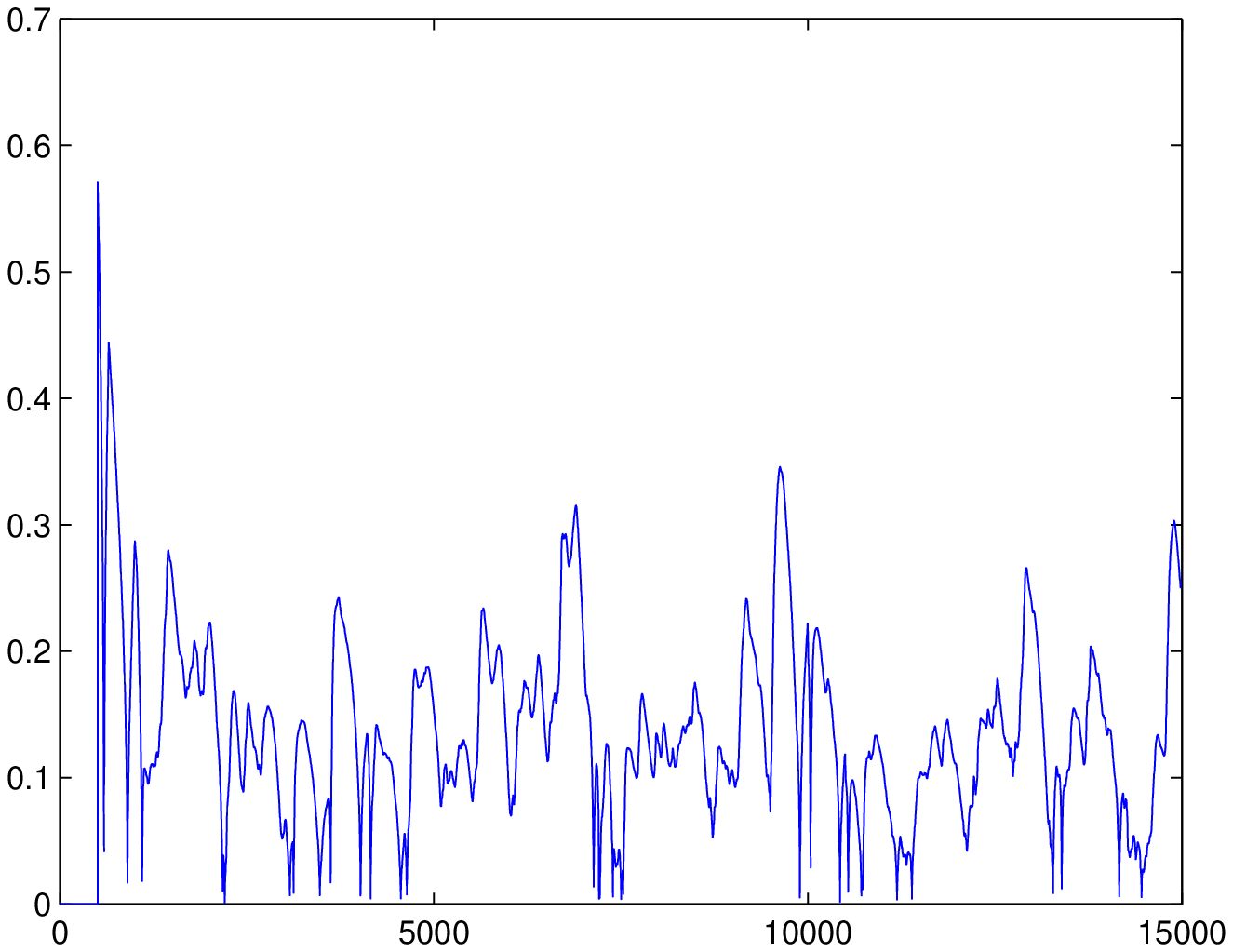}}}
\subfigure[IBM]{\rotatebox{-0}{\includegraphics*[width=.32\columnwidth]{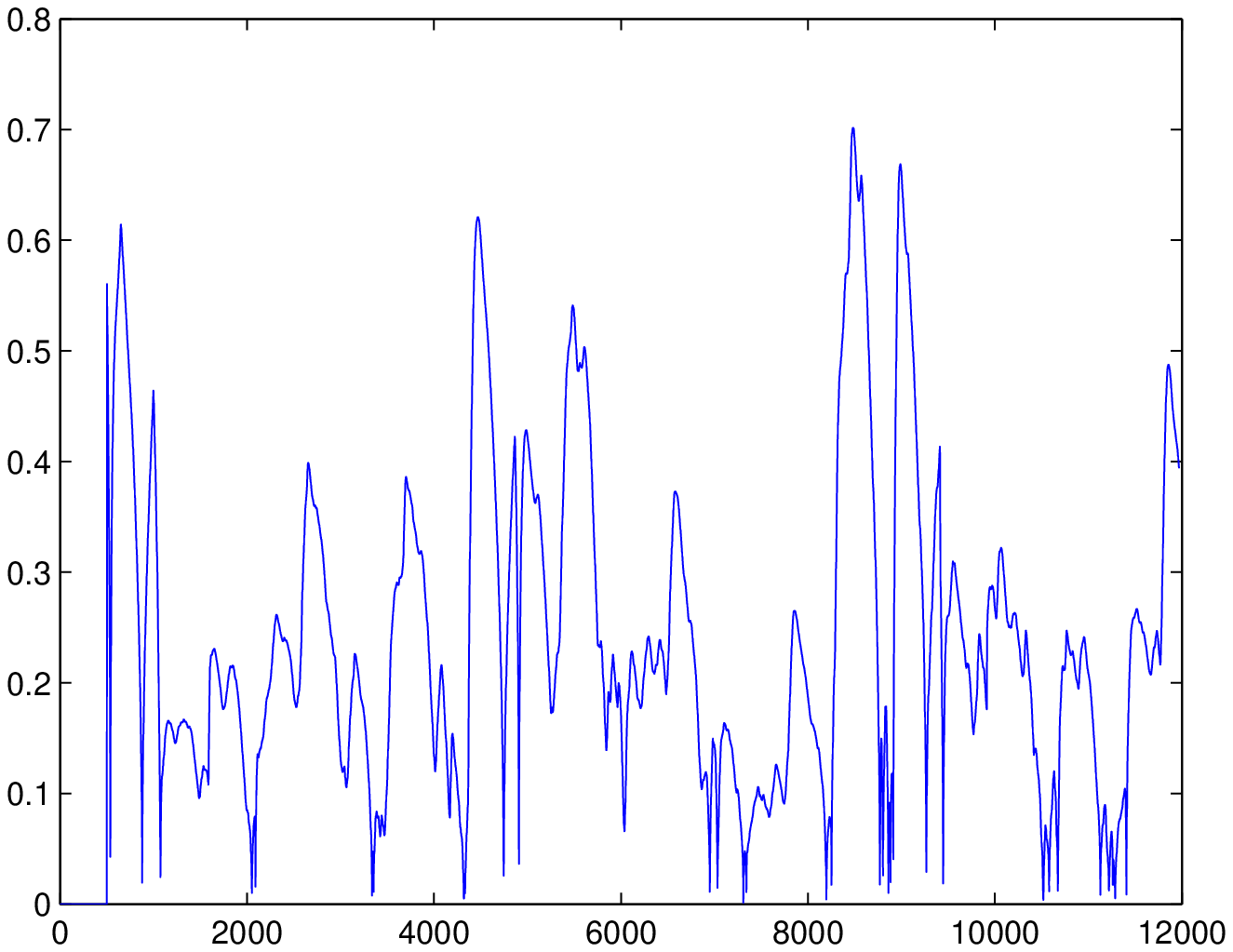}}}
\subfigure[JPM]{\rotatebox{-0}{\includegraphics*[width=.32\columnwidth]{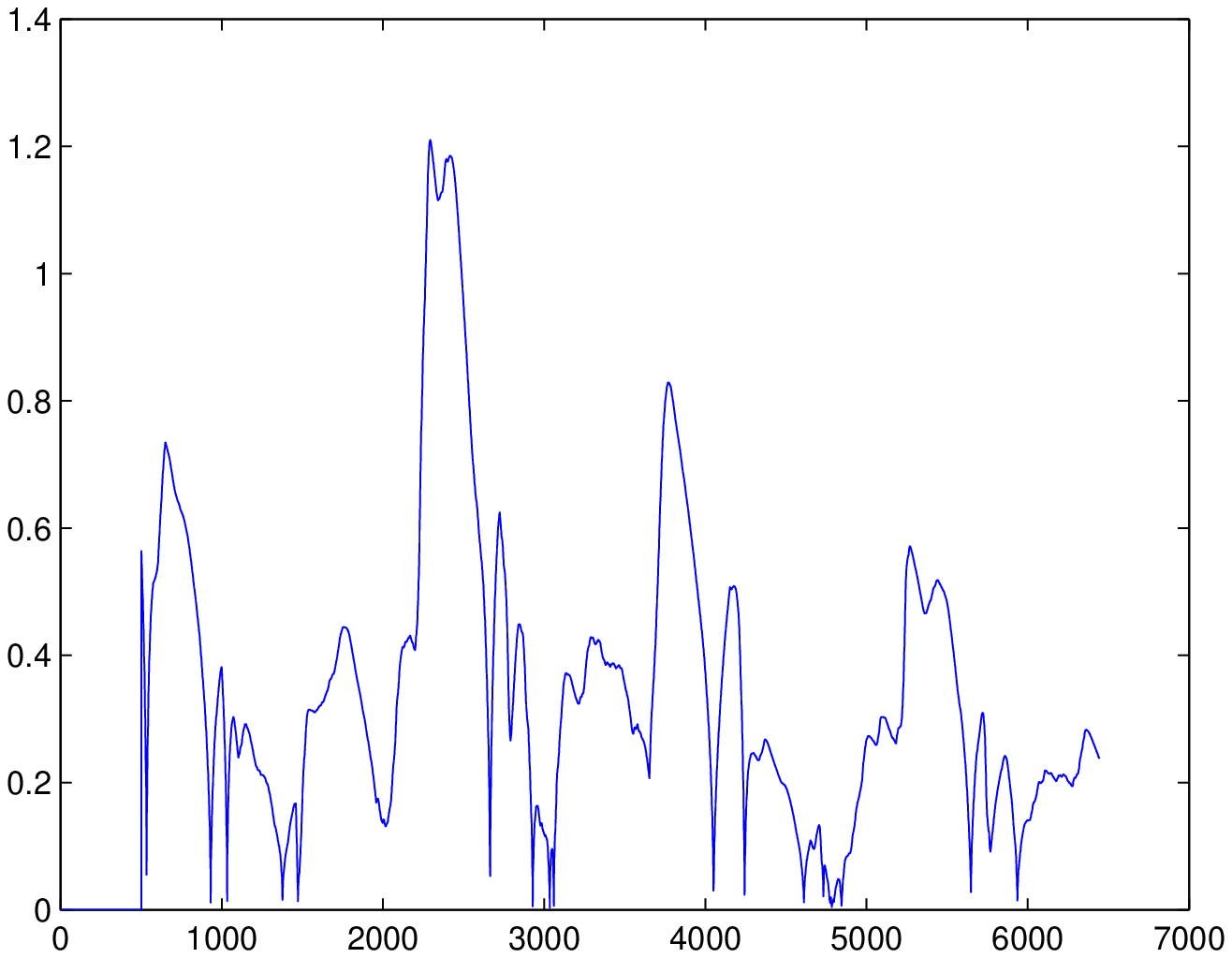}}}
\vspace{-.4cm}\caption{Volatility of returns\label{VReturn}}
\end{figure}

Figures \ref{Value}-(a), \ref{Value}-(b), \ref{Value}-(c) exhibit the
daily time series behaviors of the  the S\&P 500 and of the two
following assets:
\begin{enumerate}
\item IBM from 1962-01-02 until 2009-07-21 (11776 days),
\item JPMORGAN CHASE (JPM) from 1983-12-30 until 2009-07-21 (6267 days).
\end{enumerate}
\noindent The corresponding returns are given in Figures
\ref{Return}-(a), \ref{Return}-(b), \ref{Return}-(c) and their
volatilities in Figures \ref{VReturn}-(a), \ref{VReturn}-(b),
\ref{VReturn}-(c). We took $L = 500$ for the length $L$ of the
sliding windows.

Compare, as in Section \ref{bet}, \textit{i.e.}, without $\alpha$,
those various assets.

The comparison between IBM and S\&P 500 utilizes Formula \eqref{1}.
Figures \ref{beta}-(a), \ref{beta}-(b) and \ref{beta}-(c) show three $\beta$s
corresponding respectively to the values, the returns and the
volatilities.

\subsection{Bivariate $\beta$}
\begin{figure}
\vspace{-.4cm}
\center
\subfigure[Trend of IBM's $\beta$]{\rotatebox{-0}{\includegraphics*[width=.32\columnwidth]{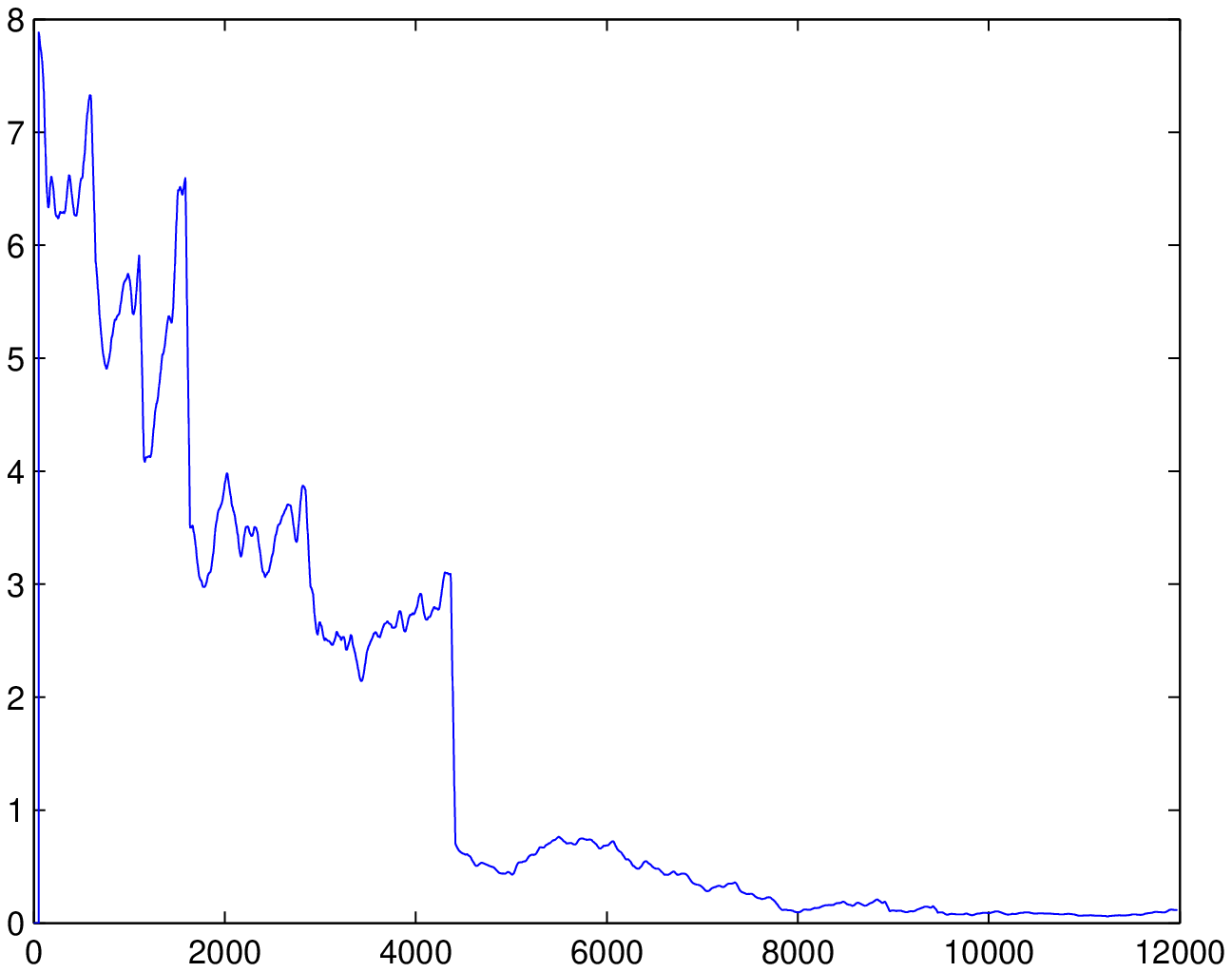}}}
\subfigure[Trend of R(IBM)'s $\beta$]{\rotatebox{-0}{\includegraphics*[width=.32\columnwidth]{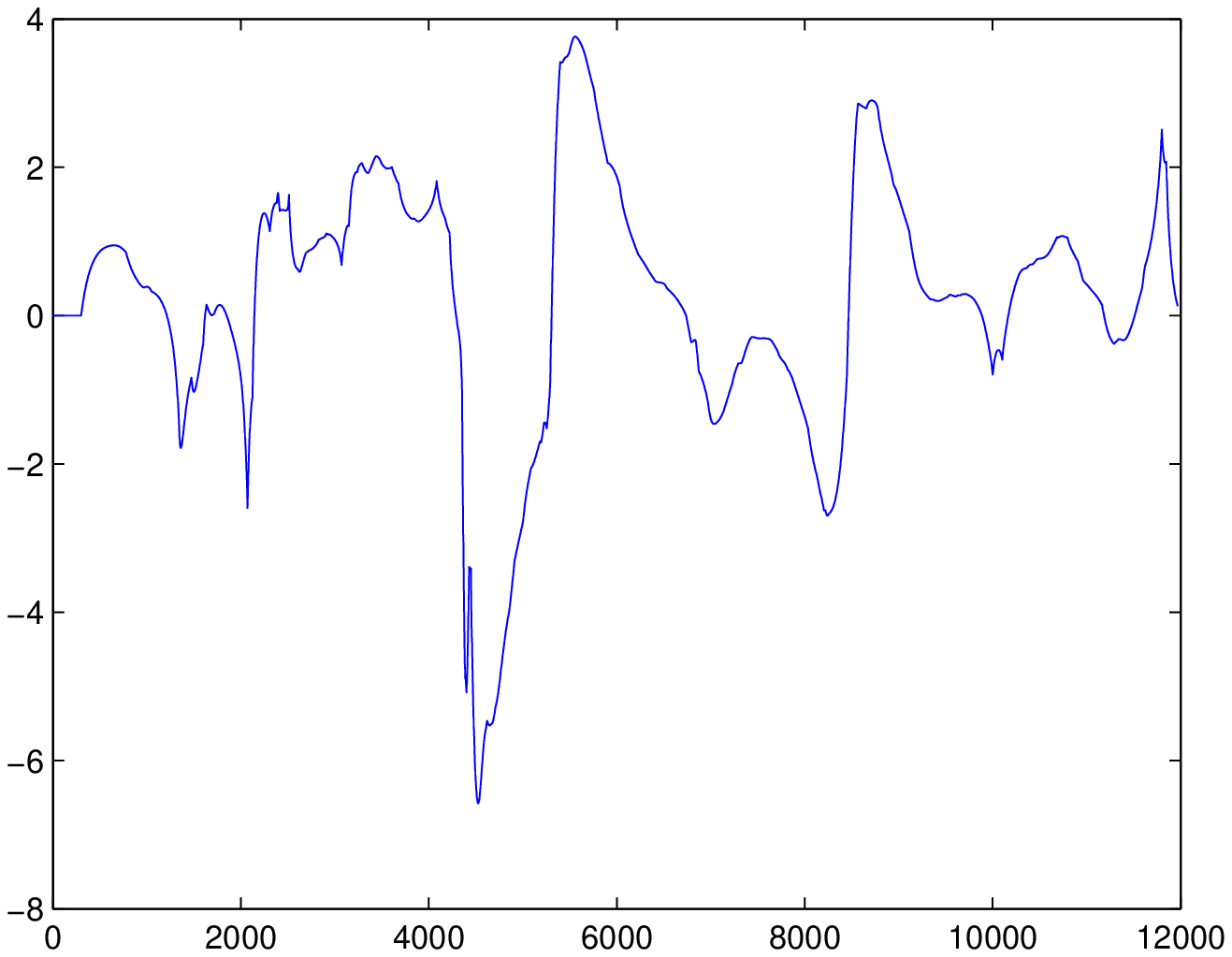}}}
\subfigure[Trend of $\beta$ of the volatility of R(IBM)]{\rotatebox{-0}{\includegraphics*[width=.32\columnwidth]{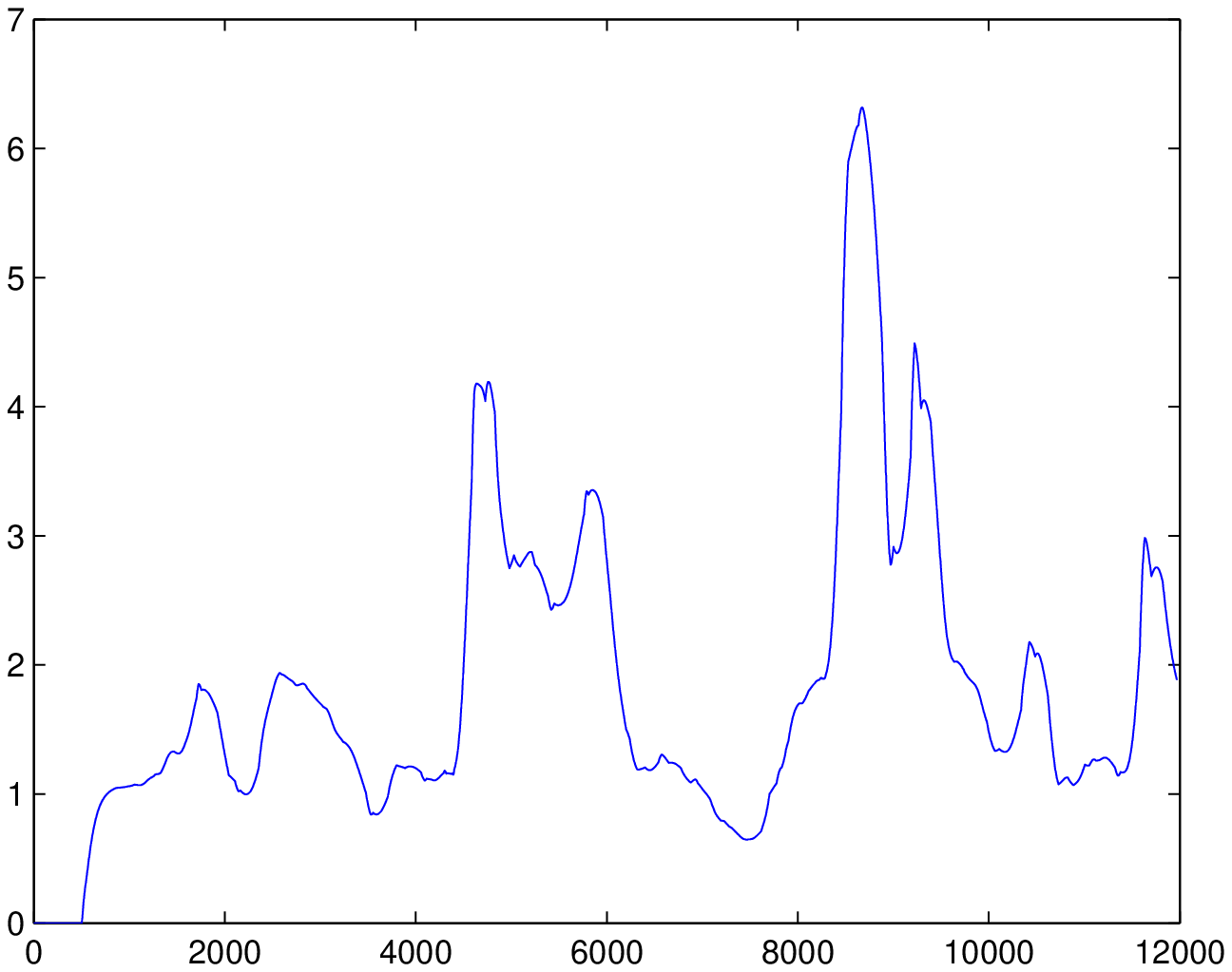}}}
\subfigure[Trends of R(IBM)'s $\beta_1$ ({\it blue}) and R(JPM)'s $\beta_2$ ({\it red})]{\rotatebox{-0}{\includegraphics*[width=.32\columnwidth]{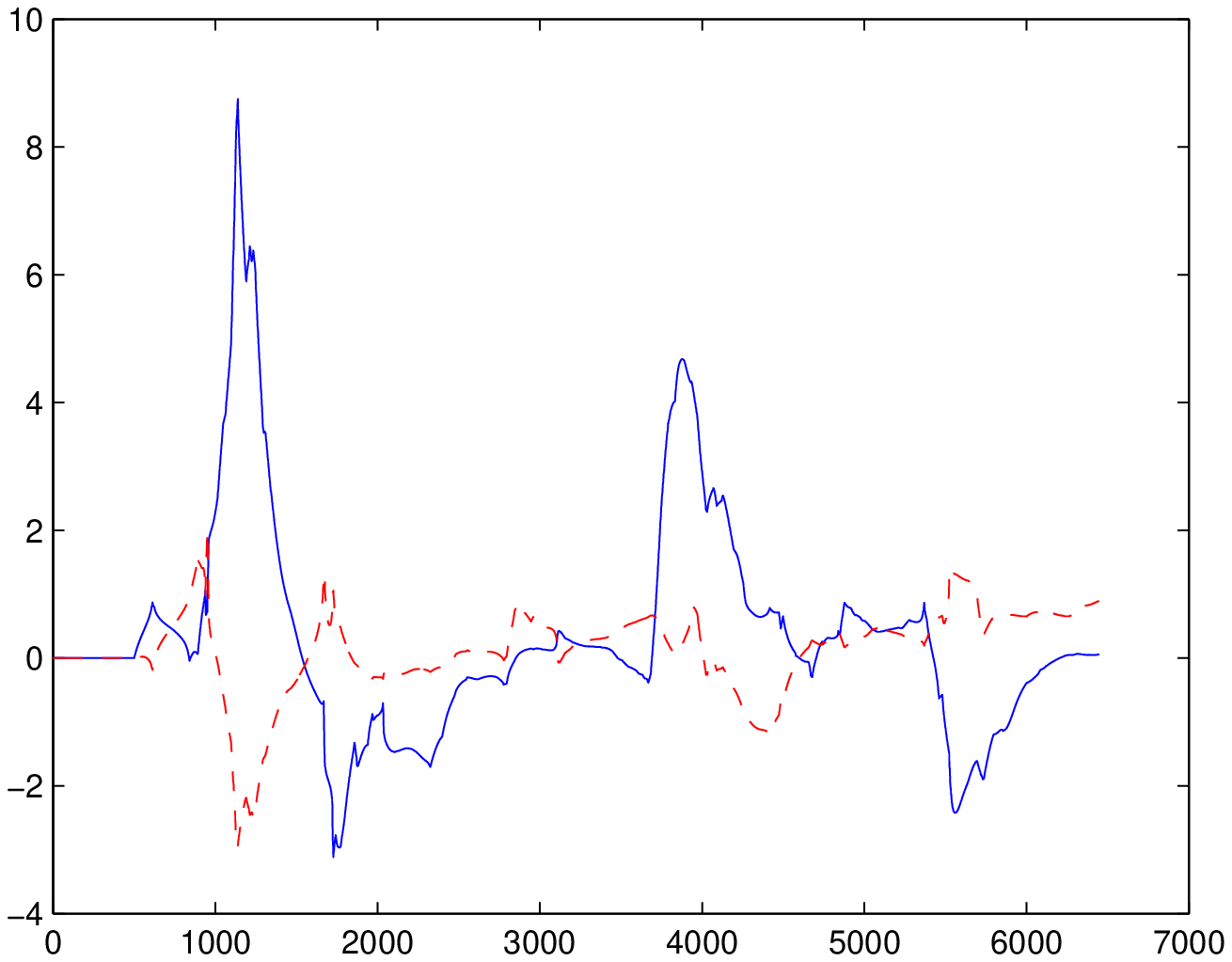}}}
\vspace{-.4cm}\caption{Betas\label{beta}}
\end{figure}

A bivariate extension is provided by a rather academic example where
we want to ``explain'' the S\&P 500 via IBM and JPM. Set therefore
$$
R_{\text{S\&P 500}} = \beta_1 R_{\text{IBM}} + \beta_2
R_{\text{JPM}}
$$
where $R_{\text{S\&P 500}}$ (resp. $R_{\text{IBM}}$,
$\text{Ret}_{\text{JPM}}$) is the return of S\&P 500 (resp. IBM,
JPM). According to Section \ref{bet} we have to invert the
determinant of the $2\times2$ matrix
{\small $$B= \begin{pmatrix} \int_{t-L}^t R_{\text{IBM}}(\tau)d\tau &
\int_{t-L}^t R_{\text{JPM}}(\tau)d\tau \\
\int_{t-L}^t\tau R_{\text{IBM}}(\tau)d\tau & \int_{t-L}^t\tau
R_{\text{JPM}}(\tau)d\tau
\end{pmatrix}$$}
Several sizes $L= 100, 300, 500$ for the sliding windows are
utilized in parallel in order, if $\det (B) \simeq 0$, to pick up
the size where $| \det (B) |$ is the greatest. Figure \ref{beta}-(d)
exhibits quite convincing results.

%\newpage

%\section{References}


\end{document}